\documentclass[12pt]{article}
\usepackage{graphicx}
\usepackage{epstopdf}
\usepackage{amsfonts}
\usepackage{amssymb}
\usepackage[footnotesize]{caption2}
\usepackage{mathrsfs}
\usepackage{amsmath}

\usepackage{authblk}

\begin{document}
\renewcommand{\thefootnote}{\fnsymbol{footnote}}
\title{Hawking radiation from the holographic screen}
\author[1]{\small Ying-Jie Zhao \thanks{E-mail address: yj\_zhao@bjut.edu.cn}}

\affil[1]{\small Institute of Theoretical Physics, Beijing University of Technology,  Beijing 100124, China}

\date{}
\maketitle
\begin{abstract}

\setlength{\parindent}{0pt} \setlength{\parskip}{1.5ex plus 0.5ex
minus 0.2ex} 
In this paper we generalize the Parikh-Wilczek scheme to a holographic screen in the framework of the ultraviolet self-complete
quantum gravity. We calculate that the tunneling probabilities of the massless and massive
particles depend on their energies of the particles and the mass of the holographic screen. The radiating temperature has not been the standard Hawking temperature.
On the contrary, the quantum unitarity principle always remains.

\vskip 10pt
\noindent
{\bf PACS Number(s)}: 04.60.-m, 04.70.Dy, 11.10.Nx
\vskip 5pt
\noindent

\end{abstract}

\thispagestyle{empty}

\newpage

\setcounter{page}{1}

According to the no-hair theorem, a black hole is generated from a celestial body's collapse can be completely characterized by only three externally observable classical parameters: mass, electric charge, and angular momentum.  In the classical Einstein gravitational theory, other information of the celestial body  is confined inside the black hole. However,
in 1974 Hawking\cite{HAWKING1,HAWKING2} recovered that a black hole is not really black previously thought but radiates energy due to the quantum effects near the event horizon, and so that the black hole has a spectrum of a black body. The pure black body spectrum  has no information and the total evaporation of the black hole give rise to the information loss paradox, which breaks the
unitarity of quantum theory.

In 1999, Parikh and Wilczek \cite{PARIKH1,PARIKH2} gave a semiclassical derivation of Hawking radiation as a tunneling process. They argued that with the continuous radiation  of particles  the energy of the black hole
decreases and the contraction of radius  of the horizon  make the particles get across the classically forbidden trajectory,  in other words, the potential barrier is created by the self-gravity of the system.
They introduced a special coordinate system where  the metric is not singular at the horizon, and then used the WKB method to compute the tunneling rate. In that article they worked out the emission rate of the massless and uncharged particles radiating from the
Schwarzchild and R-N black holes, and calculated the Hawking temperature by comparing the exponential part of the emission rate with a Boltzmann factor when neglecting the quadratic term of the energy.

In recent years, the Parikh-Wilczek method has been generalized to the more complex cases. One case is that particles have mass or charge, or both, another case is that the spacetime is no longer so simple. For example, the noncommutativity idea \cite{SEIBERG} originating from the ultraviolet divergence elimination in loop quantum theory\cite{SNYDER} has been introduced and a great many papers has been published\cite{NCPAPERS}. All the noncommutative spacetimes predict the existence of a minimal length of the order of the Planck scale. Unfortunately, many sorts of spacetimes constructed depend on external noncommutativity parameters.

In 2012, P. Nicolini and E. Spallucci\cite{NICOLINISSS1,NICOLINISS2} derived a static, neutral, non-rotating black hole metric whose extremal configuration radius is  equal to the Planck length in order to avoid introducing an additional principle to justify the existence of  a minimal length to provide a UV cut off.  Below the sub-Planckian scale the interior of the black hole loses any physical meaning thus no singularity in the origin. The authors named the particular black hole
as \emph{holographic screen}. Moreover, they discussed the thermodynamics of the holographic screen and pointed out that the area law is corrected by a logarithmic term and a minimal holographic screen corresponding to
the zero entropy existed.

Our aim is to generalize the Parikh-Wilczek method to the tunneling process of massless and uncharged particles from  holographic screen. Our article is arranged as follows. In Sections 2, a brief introduction about holographic screen is given. In section 3, we work out the tunneling rate of the massless  particles tunneling through the horizon of the holographic screen. In addition, we derive the temperature of the holographic screen and study the extremal case when its mass are very large. In section 4, we discuss the massive particles tunneling through the horizon of the holographic screen. Finally in section 5, we present a conclusion.
\\
\\
\\
\\
\section{A self-regular holographic screen}
The mass density of a point particle with mass $M$ in spherical coordinates is proportional to the Delta function
\begin{eqnarray}
\rho(r) = \frac{M}{4 \pi r^2} \delta(r),
\end{eqnarray}
which can be generalized to a derivative of a smooth function $h(r)$ \cite{NICOLINISSS}
\begin{eqnarray}
\rho(r) = \frac{M}{4 \pi r^2} \frac{d}{dr} h(r) \equiv T_0^0
\end{eqnarray}
to overcome the problem that at  the sub-Planckian energy regime the Compton wave length of a particle is larger than  a Schwarzschild  black hole  with the same mass. Taking the conservation equation $\nabla_\mu T^{\mu\nu} = 0$ into consideration the stress tensor
takes the form
\begin{eqnarray}
T_\mu^\nu = {\rm diag}\left( -\rho, p_r, p_r + \frac{r}{2} \partial_r p_r, p_r + \frac{r}{2} \partial_r p_r \right) \label{TTT}
\end{eqnarray}
with $p_r = -\rho$.
By substituting eq.(\ref{TTT}) into Einstein equation and assuming that the form of the left hand side of the Einstein equation remains unchanged
we get the metric $( {\rm gravitational \ constant}\ G= L_p^2)$
\begin{eqnarray}
ds^2 = -\left(1- \frac{2 L_p^2 m(r)}{r}\right) dt^2  + \left(1- \frac{2 L_p^2 m(r)}{r}   \right)^{-1} dr^2 +r^2 d\Omega^2,
\end{eqnarray}
where the parameter $m(r)$ takes the form
\begin{eqnarray}
m(r) =  4 \pi \int_0^r dr' {r'}^2 \rho(r').
\end{eqnarray}
The particular form of the function $h(r)$ must satisfy two rules\cite{NICOLINISSS}:
i). Spacetime in the sub-Planckian regime has no physical meaning;
ii). The characteristic scale of the system is provided by the spacetime itself of the scale of the Planck length, not imposed as a external parameter.
The most natural and algebraically assumption can be written as
\begin{eqnarray}
h(r) =  \frac{r^2}{r^2 + L_p^2},
\end{eqnarray}
where $L_p$ is the Planckian length,  and thus the smeared energy density $\rho(r)$  is
\begin{eqnarray}
\rho(r) = \frac{M L_P^2}{2 \pi r \left( r^2 + L_p^2 \right)^2}.
\end{eqnarray}

Now we have found the modified metric of holographic screen
\begin{eqnarray}
ds^2 &=& -\left(1- \frac{2L_p^2 m(r)}{r}\right) dt^2  + \left(1- \frac{2L_p^2 m(r)}{r}   \right)^{-1} dr^2 +r^2 d\Omega^2   \nonumber \\
&=& -\left( 1- \frac{2M {L_P}^2 r}{r^2+{L_p}^2} \right)dt^2 + \left( 1- \frac{2M {L_P}^2 r}{r^2+{L_p}^2} \right)^{-1}d{r^2} +r^2 d\Omega^2.
\end{eqnarray}
\\
\\
\\
\section{The Parikh-Wilczek Tunneling Mechanism and massless particles}
Now we investigate the quantum tunneling through the holographic screen via Parikh-Wilczek Tunneling mechanism.
To study the quantum tunneling through the holographic screen we considering the modified metric of the holographic screen
\begin{eqnarray}
ds^2 = -\left( 1- \frac{2M {L_P}^2 r}{r^2+{L_p}^2} \right)d{t_s}^2 + \left( 1- \frac{2M {L_P}^2 r}{r^2+{L_p}^2} \right)^{-1}d{r^2} +r^2 d\Omega^2.
\end{eqnarray}
The holographic screen admits two horizons provided $M > L_p$
\begin{eqnarray}
r_{\pm} = M L_p^2 \pm L_p \sqrt{M^2 L_p^2 -1}
\end{eqnarray}
 determined by
\begin{eqnarray}
1- \frac{2M {L_P}^2 r_{\pm}}{r^2+{L_p}^2} =0.
\end{eqnarray}
At the beginning, it is necessary to choose the Painlev\'{e} coordinate that having no singularity at the horizon.
The suitable choice can be written as \cite{PAINLEVE}
\begin{eqnarray}
dt_s = dt \pm \frac{\sqrt{(r^2+{L_p}^2)(r^2+{L_p}^2-2 M {L_p}^2 r)}}{r^2+{L_p}^2-2 M {L_p}^2 r}dr.
\end{eqnarray}
where $t$ is the  Painlev\'{e} time. After the above transformation the  Painlev\'{e} line element reads
\begin{eqnarray}
ds^2 = - \left(1 - \frac{2 M {L_p}^2 r}{r^2+{L_p}^2} \right) dt^2 + 2 \sqrt{\frac{2 M {L_p}^2 r}{r^2+{L_p}^2}}dtdr + dr^2 +r^2 d\Omega^2.
\end{eqnarray}
The radial null geodesics are calculated as
\begin{eqnarray}
\dot{r} = \frac{dr}{dt}= \pm 1 - \sqrt{ \frac{2 M L_p^2 r}{r^2+L_p^2} },
\end{eqnarray}
where the plus (minus) sign corresponding to outgoing (ingoing) geodesics.

Now we consider a massless particle radiating from the holographic screen as a massless shell through its horizon.
In the Parikh-Wilczek tunneling mechanism, the effect of self-gravitation the particles tunnel out of a holographic screen and its
energy decreases because of the total energy conversation, which makes the mass of the holographic screen decline and the horizon of the black hole shrink smaller. Naturally, the metric of the holographic screen must be edited. Here we fix the total mass $M$ and denote the energy evaporating from the holographic screen as $\omega$ and use $M \rightarrow M-\omega$, the spacetime metric can be written as
\begin{eqnarray}
ds^2 = - \left[1 - \frac{2 (M-\omega) {L_p}^2 r}{r^2+{L_p}^2} \right] dt^2 + 2 \sqrt{\frac{2 (M- \omega) {L_p}^2 r}{r^2+{L_p}^2}}dtdr + dr^2 +r^2 d\Omega^2,
\end{eqnarray}
and in the same way, the radial null geodesics are calculated as
\begin{eqnarray}
\dot{r} = \frac{dr}{dt}= \pm 1 - \sqrt{ \frac{2 (M - \omega) L_p^2 r}{r^2+L_p^2} }.
\end{eqnarray}
The characteristic length of  the massless particle  described as a spherically symmetric massless shell  is infinitesimal near the horizon on account of the infinite blue-shift, hence the wavenumber inclines to infinity.
In accordance with the WKB method, the imaginary part of the action that an s-wave massless particle  traveling on the radial null geodesics tunnel across the outer horizon $r_+$ from  the initial  position $r_{in} = M L_p^2 + L_p \sqrt{M^2 L_p^2 - 1}$ to  the final position $r_{out}=(M- \omega) L_p^2 + L_p \sqrt{(M-\omega)^2 L_p^2 - 1}$ can be exhibited as follows
\begin{eqnarray}
{\rm Im}{\cal I} = {\rm Im} \int_{r_{in}}^{r_{out}} p_r dr
 = {\rm Im} \int_{r_{in}}^{r_{out}} \int_{0}^{p_r} d{p_r}' dr
\end{eqnarray}
with the Hamilton equation
\begin{eqnarray}
d{p_r}' = \frac{dH}{\dot{r}},
\end{eqnarray}
and $H= M-\omega'$. We can deform the contour of the $r$ integral around the pole at the horizon in order to ensure the positive energy solutions decay in time (choose the lower half $\omega'$  plane), so the imaginary part of the action is worked out as
\begin{eqnarray}
{\rm Im}{\cal I} &=& - {\rm Im} \int_{0}^{\omega}d\omega' \int_{r_{in}}^{r_{out}} \frac{dr}{1-\sqrt{\frac{2 (M -\omega') L_p^2 r}{r^2+L_p^2}}}    \nonumber\\
&=& \pi \int_{0}^{\omega} \frac{  2(M-\omega') L_p^2 \left[(M-\omega') L_p + \sqrt{(M-\omega')^2 L_p^2 - 1}\right]  }{\sqrt{(M-\omega')^2 L_p^2 - 1}  }d\omega'    \nonumber  \\
&=&  \pi \omega \left(2M - \omega \right)L_p^2 + \pi L_p M \sqrt{M^2 L_p^2 -1} -\pi L_p (M-\omega) \sqrt{(M - \omega)^2 L_p^2 -1}  \nonumber \\
& &+  \pi \ln {\frac{M L_p + \sqrt{M^2 L_p^2 -1}}{(M - \omega)L_p + \sqrt{(M - \omega)^2 L_p^2 -1 }}}
\end{eqnarray}
The tunneling rate is
\begin{eqnarray}
\Gamma  &\thicksim& \exp(-2 {\rm Im}{\cal I}).
 \end{eqnarray}
Expanding $\Gamma$ with respect $\omega$
\begin{eqnarray}
\Gamma  \approx   e^{-4\pi \omega \left[ L_p^2 M+\frac{ L_p^3 M^2}{\left(L_p^2 M^2-1\right)^{1/2}}\right]-2\pi  \omega^2 \left[-L_p^2-\frac{L_p^5 M^3}{\left(L_p^2 M^2-1\right)^{3/2}}+\frac{2 L_p^3 M}{\left(L_p^2 M^2-1\right)^{3/2}}\right]},
\end{eqnarray}
and comparing the first order of $\omega$ with the Boltzmann factor $e^{-\frac{\omega}{T}}$, we obtain the temperature of the holographic screen
\begin{eqnarray}
T=\frac{\sqrt{M^2 L_p^2 - 1}}{4 \pi M L_p^2 \left(\sqrt{M^2 L_p^2 - 1}+M L_p \right)},
\end{eqnarray}
which  demonstrates the temperature of the holographic screen is no longer equivalent to the Schwartzschild metric's.
We depict the the curves of the radiation temperature $T$ versus the mass $M$ in Figure 1 for comparisons.
\begin{figure}[htbp]
\begin{center}
\includegraphics{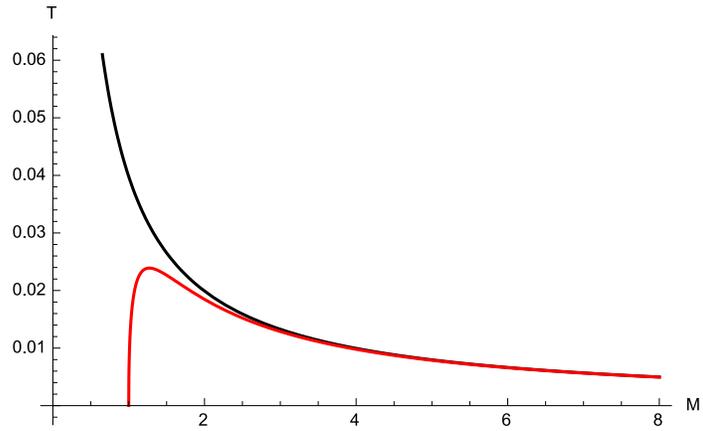}
\caption{The  radiation temperature $T$ versus the mass $M$. The black curve stands for the temperature of Schwartzschild black hole, and  the red
curve stands for the temperature of holographic screen.  Here we use the Planck units. }
\label{default}
\end{center}
\end{figure}
The entropy has a logarithmic correction,
\begin{eqnarray}
S = \int_{L_p^{-1}}^{M} \frac{dM}{T}= 2\pi  \left[ L_p^2 M^2-1 +L_p M \sqrt{L_p^2 M^2 -1}+ \ln\left( L_p M + \sqrt{ L_p^2 M^2 -1}\right)\right].
\end{eqnarray}
The difference of the entropy after and before the radiation
\begin{eqnarray}
\Delta S &=& S(M-\omega) - S(M)   \nonumber \\
&=& 2\pi  \left[  (\omega  - 2 M)\omega L_p^2 +  (M-\omega)L_p \sqrt{ (M-\omega)^2 L_p^2 -1}- M L_p \sqrt{L_p^2 M^2 -1}\right. \nonumber \\
& &\left. + \ln\frac{ (M-\omega) L_p  + \sqrt{(M-\omega)^2 L_p^2 -1}}{  M L_p + \sqrt{  M^2 L_p^2  -1}}\right].
\end{eqnarray}
We depict the the curves of the entropy $S$ versus the mass $M$ in Figure 2 for comparisons.
\begin{figure}[htbp]
\begin{center}
\includegraphics{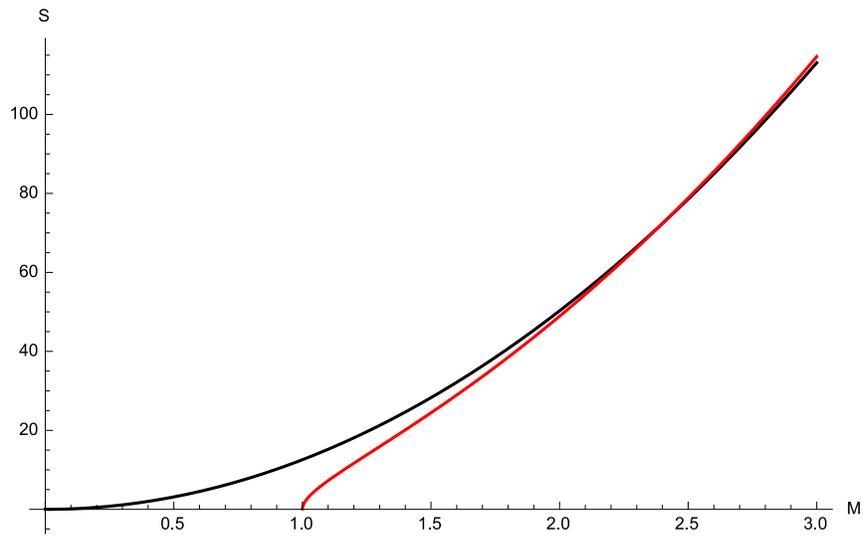}
\caption{The  entropy $S$ versus the mass $M$. The black curve stands for the entropy of Schwartzschild black hole, and  the red curve stands for the entropy of holographic screen.  Here we use the Planck units. }
\label{default}
\end{center}
\end{figure}
$M \gg{L_p^{-1}}$, the action can be simplified as
\begin{eqnarray}
{\rm Im}{\cal I} \approx 2\pi \omega \left(2M - \omega \right)L_p^2 +\pi \ln \frac{M}{M-\omega}.
\end{eqnarray}
And the tunneling rate is
\begin{eqnarray}
\Gamma  \thicksim e^{-8\pi M L_p^2 \omega -4 \pi \omega^2(2/M^2-L_p^2) }.
\end{eqnarray}
The temperature degenerates to the conventional results of the Schwartzschild black hole,
\begin{eqnarray}
T \approx \frac{1}{8 \pi M L_p^2  },
\end{eqnarray}
while the entropy reduces to
\begin{eqnarray}
S \approx 4\pi   L_p^2 M^2+ 2 \pi\ln (2 L_p M).
\end{eqnarray}
\\
\\
\\
\section{Massive particles}

In the section  we discuss  massive particles radiating from the holographic screen as a shell through its horizon. A massive particle
because of the particles have   which makes the mass of the holographic screen decline and the horizon of the black hole shrink smaller.  A massive particle
has no longer moved along a null geodesics  but  a time-like trajectory determined by the  Lagrangian,
\begin{eqnarray}
{{\cal L}(x^\mu, \tau)} = \frac{m}{2} g_{\mu\nu} \frac{dx^\mu}{d\tau} \frac{dx^\nu}{d\tau},
\end{eqnarray}
where $\tau$ stands for the proper time.
\begin{eqnarray}
ds^2 = - \left(1 - \frac{2 M {L_p}^2 r}{r^2+{L_p}^2} \right) dt^2 + 2 \sqrt{\frac{2 M {L_p}^2 r}{r^2+{L_p}^2}}dtdr + dr^2 +r^2 d\Omega^2.
\end{eqnarray}
The fact that the Lagrangian does not contain  canonical coordinate $t$  implies
in  its corresponding Euler-Lagrange equation the corresponding canonical momentum of $t$, also the particle's energy is conserved ($\dot{t} = \frac{dt}{d\tau}, \dot{r} = \frac{dr}{d\tau}$),
\begin{eqnarray}
 p_{t} = \frac{\partial{\cal L}}{\partial \dot{t}} =m\left[  -\left(1 - \frac{2 M {L_p}^2 r}{r^2+{L_p}^2} \right) {\dot t_p} +\sqrt{\frac{2 M {L_p}^2 r}{r^2+{L_p}^2}}{\dot r}\right]= -\omega,   \label{PTP}
\end{eqnarray}
here  the minus sign before $\omega$ results from  the positivity of the energy  of the tunneling particle.
For a time-like trajectory, we also have
\begin{eqnarray}
  g_{\mu\nu} \frac{dx^\mu}{d\tau} \frac{dx^\nu}{d\tau} = -1,
\end{eqnarray}
i.e.,
\begin{eqnarray}
- \left(1 - \frac{2 M {L_p}^2 r}{r^2+{L_p}^2} \right){\dot t}^2 +2 \sqrt{\frac{2 M {L_p}^2 r}{r^2+{L_p}^2}}{\dot t}{\dot r} +{\dot r}^2 =   -1,     \label{GUV-1}
\end{eqnarray}
and in consequence of Eq.(\ref{PTP}) and (\ref{GUV-1})  the  particle's trajectory  along
the radial direction is

\begin{eqnarray}
\frac{dr}{dt} = \frac{- g_{00}  \sqrt{g_{00}m^2 +\omega^2} }{\omega\sqrt{g_{01}^2-g_{00}}+ g_{01}\sqrt{g_{00} m^2+\omega^2}}.
\end{eqnarray}
where
\begin{eqnarray}
g_{00}=- \left(1 - \frac{2 M {L_p}^2 r}{r^2+{L_p}^2} \right), \ \ \ \ \ \ \ \ g_{01}=\sqrt{\frac{2 M {L_p}^2 r}{r^2+{L_p}^2}}.
\end{eqnarray}
By repeating the same steps in Section 3 , the imaginary part of the action that a massive particle  along the radial direction tunnels across the outer horizon $r_+$ from $r_{in} = M L_p^2 + L_p \sqrt{ M^2 L_p^2 - 1}$ to $r_{out}=(M- \omega) L_p^2 + L_p \sqrt{(M-\omega)^2 L_p^2 - 1}$ can be exhibited as follows
\begin{eqnarray}
{\rm Im}{\cal I}  &=&  {\rm Im} \int_{r_{in}}^{r_{out}} \int_{m}^{\omega'} \frac{dH}{\dot{r}} dr               \nonumber\\
&=& -{\rm Im} \int_{m}^{\omega}d\omega' {\int_{r_{in}}^{r_{out}} \frac{\left(\omega'\sqrt{g_{01}^2-g_{00}}+ g_{01}\sqrt{g_{00} m^2+\omega'^2}\right)  dr}{- g_{00}  \sqrt{g_{00}m^2 +\omega'^2}}  }  \nonumber\\
&=& - {\rm Im} \int_{m}^{\omega}d\omega' {\int_{r_{in}}^{r_{out}} \frac{(r^2+L_p^2)\left(\omega'\sqrt{g_{01}^2-g_{00}}+ g_{01}\sqrt{g_{00} m^2+\omega'^2}\right)  dr}{(r - r_+)(r-r_-)  \sqrt{g_{00}m^2 +\omega'^2}}  },
\end{eqnarray}
where the two horizons are
\begin{eqnarray}
r_\pm = (M- \omega') L_p^2 \pm L_p \sqrt{(M-\omega')^2 L_p^2 - 1}, \ \ \ \ \ \ \  H= M-\omega'.
\end{eqnarray}
Clearly, there exists a pole at $ r= (M- \omega') L_p^2 + L_p \sqrt{(M-\omega')^2 L_p^2 - 1} $, and hence  we can deform the contour of the $r$ integral around the pole at the horizon in order to ensure the positive energy solutions decay in time (choose the lower half $\omega'$  plane), so the imaginary part of the action is expressed as
\begin{eqnarray}
{\rm Im}{\cal I}  &=&  2\pi \int_{0}^{\omega}  \frac{r_+^2 + L_p^2}{r_+ - r_-} d\omega'     \nonumber \\
& =&  \pi M^2 L_p^2 + \pi L_p M-\sqrt{ M^2 L_p^2-1}   + \pi \ln \left[M L_p +\sqrt{M^2 L_p^2-1}\right]                              \nonumber \\
&&- \pi (M-\omega)^2L_p^2 - \pi L_p(M-\omega)\sqrt{(M-\omega)^2 L_p^2-1}       \nonumber \\
&&- \pi \ln \left[(M-\omega)L_p +\sqrt{(M-\omega)^2 L_p^2-1}\right]
\end{eqnarray}
The tunneling rate is
\begin{eqnarray}
\Gamma  &\thicksim& \exp(-2 {\rm Im}{\cal I}).
 \end{eqnarray}
Expanding $\Gamma$ with respect $\omega$
\begin{eqnarray}
\Gamma  &\approx&   e^{-4 \pi  L^2 M   \left(\frac{L M}{\sqrt{L^2 M^2-1}}+1\right) \omega}    \nonumber\\
&&e^{\frac{2 \pi  L^2 \left(4 L^5 M^5-9 L^3 M^3-7 L^2 M^2 \sqrt{L^2 M^2-1}+\sqrt{L^2 M^2-1}+4 L^4 M^4 \sqrt{L^2 M^2-1}+4 L M\right)}{\left(L^2 M^2-1\right)^{3/2} \left(\sqrt{L^2 M^2-1}+L M\right)^2}\omega ^2 }
\end{eqnarray}
and comparing the first order of $\omega$ with the Boltzmann factor $e^{-\frac{\omega }{T}}$, we obtain the temperature of the holographic screen
\begin{eqnarray}
T=\frac{\sqrt{M^2 L_p^2 - 1}}{4 \pi M L_p^2 \left(\sqrt{ M^2 L_p^2 - 1}+M  L_p \right)},
\end{eqnarray}
which  demonstrates the temperature of the holographic screen is no longer equivalent to the Schwartzschild metric's.
\\
\\
\\
\section{Conclusion}
In this paper, we have applied the the Parikh-Wilczek scheme on a holographic screen, analyzed the tunneling process of the massless and massive particles and derived their tunneling rates. We have pointed out that the radiation spectra are not purely thermal,
and the temperatures of the holographic screen are not equal to the standard Hawking temperature of the Schwarzschild black hole. Moreover, we have also noticed that the changes of the entropies $\Delta S = -2{\rm Im}{\cal I}$  thus the unitarity principle
remains in the tunneling process of the holographic screen model.

In our future work the approach in this article will be developed and massive charged particles emitting from
the holographic screen will be studied. It is expected that several interesting results will be gained.


\begin{thebibliography}{99}
\bibitem{HAWKING1} S.W. Hawking, {\em Black hole explosions}, Nature {\bf 248}, 30 (1974).
\bibitem{HAWKING2} S.W. Hawking, {\em Particle creation by black holes}, Commun. Math. Phys. {\bf 43}, 199 (1975) (Erratum-ibid. {\bf 46}, 206 (1976)).
\bibitem{PARIKH1} M.K. Parikh and F.Wilczek, {\em Hawking radiation as tunneling}, Phys. Rev. Lett. {\bf 85} (2000)
5042 [arXiv:hep-th/9907001].
\bibitem{PARIKH2}M.K. Parikh, {\em A secret tunnel through the horizon}, Int. J. Mod. Phys. D {\bf 13} (2004) 2351;
Gen. Rel. Grav. {\bf 36} (2004) 2419 [arXiv:hep-th/0405160].
\bibitem{SEIBERG}N. Seiberg and E. Witten, {\em String theory and noncommutative geometry}, JHEP {\bf 09} (1999)
032 [arXiv:hep-th/9908142].
\bibitem{SNYDER}H.S. Snyder, {\em Quantized space-time}, Phys. Rev. {\bf 71} (1947) 38; {\em The electromagnetic field
in quantized spacetime}, Phys. Rev. {\bf 72} (1947) 68.
\bibitem{NCPAPERS} M.R. Douglas and N.A. Nekrasov, Noncommutative field theory, Rev. Mod. Phys. {\bf 73}
(2001) 977 [arXiv:hep-th/0106048]; \\
R.J. Szabo, {\em Quantum field theory on noncommutative spaces}, Phys. Rept. {\bf 378} (2003)
207 [arXiv:hep-th/0109162]; \\
R.J. Szabo, {\em Symmetry, gravity and noncommutativity}, Class. Quant. Grav.{\bf 23} (2006)
R199 [arXiv:hep-th/0606233];\\
S. Capozziello, G. Lambiase, and G. Scarpetta, {\em The generalized uncertainty principle from
quantum geometry}, Int. J. Theor. Phys. {\bf 39}, 15 (2000); \\
  F. Scardigli, {\em Generalized uncertainty principle in quantum gravity from micro-black hole
gedanken experiment}, Phys. Lett. B  {\bf  452}, 39 (1999) [arXiv:hep-th/9904025];\\
A. Kempf, G.Mangano, and R.B. Mann, {\em Hilbert space representation of the minimal length
uncertainty relation}, Phys. Rev. D {\bf 52}, {\bf 1108} (1995) [arXiv:hep-th/9412167];\\
M. Maggiore, {\em A generalized uncertainty principle in quantum gravity}, Phys. Lett. B {\bf 304},
65 (1993) [arXiv:hep-th/9301067];\\
M. Maggiore, {\em The algebraic structure of the generalized uncertainty principle}, Phys. Lett.
B  {\bf 319}, 83 (1993) [arXiv:hep-th/9309034];\\
M. Maggiore, {\em Quantum groups, gravity, and the generalized uncertainty principle}, Phys.
Lett. B  {\bf 49}, 5182 (1994) [arXiv:hep-th/9305163];\\
R.J. Adler, P. Chen, and D.I. Santiago,{\em The generalized uncertainty principle and black
hole remnants}, Gen. Rel. Grav.  {\bf 33}, 2101 (2001) [arXiv:gr-qc/0106080];\\
K. Nozari and A.S. Sefiedgar, {\em Comparison of approaches to quantum correction of black
hole thermodynamics}, Phys. Lett. B {\bf 635}, 156 (2006) [arXiv:gr-qc/0601116];\\
K. Nozari and A.S. Sefiedgar, {\em On the existence of the logarithmic correction term in black
hole entropy-area relation}, Gen. Rel. Grav.  {\bf 39}, 501 (2007) [arXiv:gr-qc/0606046];\\
W. Kim, E.J. Son, and M. Yoon, {\em Thermodynamics of a black hole based on a generalized
uncertainty principle}, JHEP  {\bf 0801}, 035 (2008) [arXiv:0711.0786 [gr-qc]];\\
K. Nozari and S.H. Mehdipour, {\em Hawking radiation as quantum tunneling from a
noncommutative Schwarzschild black hole}, Class. Quantum Grav.  {\bf 25}, 175015 (2008)
[arXiv:0801.4074 [gr-qc]];\\
K. Nozari and S.H. Mehdipour, {\em Quantum gravity and recovery of information in black hole
evaporation}, Europhys. Lett.  {\bf 84}, 20008 (2008) [arXiv:0804.4221 [gr-qc]];\\
L. Xiang and X.Q. Wen, {\em A heuristic analysis of black hole thermodynamics with general-
ized uncertainty principle}, JHEP {\bf 0910}, 046 (2009) [arXiv:0901.0603 [gr-qc]];\\
K. Nozari and S.H. Mehdipour, {\em Parikh-Wilczek tunneling from noncommutative higher
dimensional black holes}, JEHP  {\bf 0903}, 061 (2009) [arXiv:0902.1945 [hep-th]];\\
Y.-G. Miao, Z. Xue and S.-J. Zhang, {\em Massive charged particle¡¯s tunneling from spherical
charged black hole}, Europhys. Lett.  {\bf 96} (2011) 10008 [arXiv:1012.0390[hep-th]].  \\
A. Smailagic and E. Spallucci, {\em Feynman path integral on the noncommutative plane}, J.
Phys. A  {\bf 36} (2003) L467 [arXiv:hep-th/0307217];  \\
A. Smailagic and E. Spallucci, {\em UV divergence-free QFT on noncommutative
plane}, J. Phys. A  {\bf 36} (2003) L517 [arXiv:hep-th/0308193]; \\
A. Smailagic and E. Spallucci, {\em Lorentz invariance, unitarity and UV-finiteness of QFT on noncommutative spacetime}, J. Phys. A  {\bf 37} (2004) 1 (Erratum-ibid. A  {\bf 37} (2004) 7169) [arXiv:hep-th/0406174];\\
P. Nicolini, A. Smailagic and E. Spallucci, {\em  Noncommutative geometry inspired
Schwarzschild black hole}, Phys. Lett. B  {\bf 632} (2006) 547 [arXiv:gr-qc/0510112]; \\
G. Amelino-Camalia, J.R. Ellis, N.E. Mavromatos, D.V. Nanopoulos and S.Sarkar, {\em  Potential Sensitivity of Gamma-Ray Burster Observations to Wave Dispersion in Vacuo}, Nature \textbf{393}, 763 (1998) [arXiv:astro-ph/9712103];\\
G. Amelino-Camalia, {\em  Improved limit on quantum-spacetime modifications of Lorentz symmetry from observations of gamma-ray blazars}, New J.Phys.{ \bf 6}:188,2004 [arXiv:gr-qc/0212002];\\
J. Magueijo and L. Smolin, {\em Gravity's rainbow}, Class. Quant. Grav. {\bf 21} (2004) 1725-1736  [arXiv:gr-qc/0305055];\\
G. Amelino-Camalia, J.R. Ellis, N.E. Mavromatos, D.V. Nanopoulos and S.Sarkar, {\em Distance measurement and wave dispersion in a Loiville string approach to quantum gravity}, Int. J. Mod. Phys. A \textbf{12} (1997) 607-624 [arXiv:hep-th/9605221];\\
G. Amelino-Camalia, {\em Quantum-Spacetime Phenomenology}, Living Rev. Rel. {\bf 16} (2013) 5 [arXiv:0806.0339];\\
G. Amelino-Camelia, {\em Relativity in space-times with short-distance structure governed by an observer-independent (Planckian) length scale},  Int. J. Mod. Phys. D {\bf 11} (2002) 35-60 [arXiv:gr-qc/0012051];\\
J. Magueijo, L. Smolin, {\em Lorentz Invariance with an Invariant Energy Scale}, Phys. Rev. Lett. {\bf 88}, 190403 (2002);\\
A. F. Ali, {\em Black hole remnant from gravity's rainbow}, Phys. Rev. D {\bf 89} (2014) 104040 [arXiv:hep-th/1402.5320]; \\
A. F. Ali, M. Faizal, M. M. Khalil, {\em Remnants of black rings from gravity's rainbow}, JEHP {\bf 1412} (2014) 159 [arXiv:hep-th/1402.5320];\\
A. F. Ali, M. Faizal, M. M. Khalil, {\em Remnants of all black objects due to gravity's rainbow}, Mod. Phys. B {\bf 894} (2015) 341-360 [arXiv:hep-th/1410.5706].
P. Pedram, {\em New approach to nonperturbative quantum mechanics with minimal length uncertainty}, Phys. Rev. D \textbf{85}, 024016 (2012) [arXiv:1112.2327 [hep-th]]; \\
P. Pedram, {\em A higher order GUP with minimal length uncertainty and maximal momentum}, Phys. Lett. B \textbf{714}, 317 (2012)  [arXiv:1110.2999 [hep-th]]; \\
P. Pedram, {\em A higher order GUP with minimal length uncertainty and maximal momentum II: Applications}, Phys. Lett. B \textbf{718}, 638 (2012) [arXiv:1210.5334 [hep-th]].
S. Das and E. C. Vagenas, {\em Universality of quantum gravity corrections}, Phys. Rev. Lett. \textbf{101}, 221301 (2008)  [arXiv:0810.5333 [hep-th]]; \\
S. Das and E.C. Vagenas, {\em Phenomenological implications of the generalized uncertainty principle}, Can. J. Phys. \textbf{87}, 233 (2009) [arXiv:0901.1768 [hep-th]].
\bibitem{NICOLINISSS1}P. Nicolini, E. Spallucci, {\em Holographic screens in vltravoilet sel-complete quantum gravity}, AHEP {\bf 2014} (2014) 805684 [arXiv:hep-th/1210.0015].
\bibitem{NICOLINISS2}A. M. Frassino, S. K\"{o}ppel and P. Nicolini, {\em Geometric model of black hole quantum $N$-portrait, extradimensions and thermodynamics}, Entropy 2016 {\bf 18} (5) 181 [arXiv:gr-qc/1604.03263].
 \bibitem{PAINLEVE}P. Painlev\'{e}, {\em La m\'{e}canique classique er la thorie de relativit\'{e}}, C. R. Acad. Sci. (Paris)
{\bf 173} (1921) 677.
\end{thebibliography}
\end{document}